\documentclass[a4paper,fleqn]{cas-sc}

\usepackage[authoryear,longnamesfirst]{natbib}
\usepackage{hyperref}
\begin{document}
\let\WriteBookmarks\relax
\def\floatpagepagefraction{1}
\def\textpagefraction{.001}
\shorttitle{}
\shortauthors{K Sharma et~al.}

\title [mode = title]{Growth and dynamics of Econophysics:  A bibliometric and network analysis }                      

%

\author[1]{Kiran Sharma} \corref{cor1}
\ead{kiransharma1187@gmail.com}
\credit{Conceived and designed the analysis; Collected the data; Contributed data or analysis tools; Performed the analysis; Writing - original draft}
\address[1]{Chemical \& Biological Engineering, Northwestern University, Evanston, Illinois-60208, USA}
\cortext[cor1]{Corresponding author}
\author[2]{ Parul Khurana}
\ead{parul11183@gmail.com }
\credit{Collected the data; Contributed data or analysis tools; Performed the analysis; Writing - review}
\address[2]{School of Computer Applications, Lovely Professional University, Phagwara, Punjab-144401, India}

\begin{abstract}
Digitization of publications, advancement in communication technology, and the availability of bibliographic data have made it easier for the researchers to study the growth and dynamics of any discipline.
We present a study on ``Econophysics'' metadata extracted from Web of Science managed by the Clarivate Analytics from 2000-2019. The study highlights the growth and dynamics of the discipline by measures of a number of publications, citations on publications, other disciplines contribution, institutions participation, country-wise spread, etc. We investigate the impact of self-citations on citations with every five-year interval. Also, we find the contribution of other disciplines by analyzing the cited references. Results emerged from micro, meso and macro-level analysis of collaborations show that the distributions among authors collaboration and affiliations of authors follow a power law. Thus, very few authors keep producing most of the papers and are from few institutions. We find that China is leading in the production of a number of authors and a number of papers; however, shares more of national collaboration rather than international, whereas the USA shares more international collaboration. Finally, we demonstrate the evolution of the author's collaborations and affiliations networks from 2000-2019. Overall the analysis reveals the ``small-world'' property of the network with average path length 5. As a consequence of our analysis, this study can serve as an in-depth knowledge to understand the growth and dynamics of Econophysics network both qualitatively and quantitatively.

\end{abstract}


\begin{highlights}
\item The objective of this research is to assess the scientific development of ``Econophysics'' in terms of the total number of studies carried out over two decades. This study analyses the growth of publications, citations, key disciplines, journals, nations, organizations, and authors involved in information productivity and dissemination of knowledge. For the analysis, the data has been taken from the Clarivate Analytics Web of Science from 2000-2019.

\item We have shown the dynamics of citations and self-citations with a gap of five-year intervals. Also, the contribution of disciplines has been measured from the cited references.

\item We have identified the disciplines from the cited references that have highly contributed to the development of the field. 

\item We have explored the scientific collaborations networks at the micro, meso, and macro-levels to investigate the key author's, their affiliations, and corresponding countries, respectively.

\item We have demonstrated the evolution and growth of co-authorship and institutional collaborations networks over the years.

\end{highlights}


\maketitle
\section{Introduction}
\label{sec:Intro}

Scientific collaborations have seen considerable growth in recent times and have emerged as an important factor for productive and qualitative research. The citation analysis of the scientific publications have become a tool to analyze the individual's performance, journal's impact as well as the discipline's growth~\citep{ zeng2017science, radicchi2017quantifying}.
Bibliometrics analysis not only makes a decision on the researcher's growth, in fact, it also measures the growth of a discipline. Many new interdisciplinary and multidisciplinary fields have arisen over time which in turn have increased and strengthened the interdisciplinary collaborations~\citep{amaral1999econophysics, stanley1999econophysics,chakraborti2016can}. One such interdisciplinary field is ``Econophysics'' which was coined by H. Eugene Stanley in 1995~\citep{stanley2000introduction, chakrabarti2006econophysics, rosser2008econophysics}. Initially, physicists and economists contributed together to start this field and started applying theories and methods of physics to address problems in economics and stock markets~\citep{ carbone2007we, roehner2010fifteen, chakraborti2011econophysics, pereira2017econophysics, abergel2019new}. Later on, with the acceptance of the idea, scholars from other disciplines started contributing. Before the term Econophysics was coined, many people from different branches of science had worked and applied their knowledge in the field of economics leading to the evolution of Econophysics~\citep{dash2014evolution}.

Citations play a significant role in understanding the link between scientific works~\citep{tahamtan2019citation}. Nowadays, most of the research publications are created by teams of researchers instead of single individuals~\citep{guimera2005team}.
To investigate the patterns and trends of scientific collaboration, researchers have been working on publications data for a long time. There are different methods available in the literature to study collaborations and among them investigating the co-authorship network is the popular one~\citep{sun2017coauthorship}.
A co-authorship network is a social network built on scientific collaborations, and thus it is amenable to social network analysis~\citep{barabasi2016network, singh2020evolution}. With the development of complex network theory, researchers have been using network science to re-investigate the structural properties of co-authorship networks ~\citep{price1965networks, newman2001scientific, newman2003structure, newman2006structure, zheleva2009co}.

Over time many such networks have been studied in different domains of social aspects like the author's collaborations~\citep{newman2001scientific, andrikopoulos2016four}, author's affiliations collaborations~\citep{zheleva2009co}, and countries collaborations networks~\citep{ortega2013institutional}. Finding communities inside network~\citep{good2010performance} and calculating centralities have been a major focus of social network analysis~\citep{Freeman_1977, valente2008correlated}.
It identifies critical pointers in the network and often used to equate popularity and leadership. The above-mentioned social networks are either directed or undirected where nodes act as authors and edges represent the collaboration among authors. The author's collaboration analysis is a micro-level study however, such interactions among authors also give rise to institutional collaborations at meso-level and cross-country collaborations at the macro-level. Investigating the co-authorships network can help to identify entrants, leading researchers, and new collaborations. Co-authorship, institutional, and cross-country collaboration networks jointly reveal scientific collaboration and its growth~\citep{chakrabarti2010fifteen, ghosh2013econophysics, sinatra2016quantifying}. This way we captured the changes in network structure at the microscopic, mesoscopic, and macroscopic levels and identified the key leaders at all levels.

The scholars have studied the econophysicists collaboration network earlier~\citep{ fan2004network, li2007econophysicists}, however, to the best of our knowledge,
no one has performed systematic empirical research highlighting the patterns in data, key disciplines by cited references, and the patterns of collaborations at micro, meso, and macro-levels. At the micro-level, an author's collaboration, at the meso-level author's affiliation, and at macro-level countries' collaboration networks have been analyzed that demonstrate the in-depth knowledge of the growth of the discipline. This is the first time we are showing the detailed analysis of Econophysics through bibliometric and network analyses which cover the gap of the previous studies accomplished on it. It demonstrates the current state of Econophysics and provides researchers and practitioners with up-to-date knowledge. Thus, the objective of this study is to appraise the scientific evolution of Econophysics through various factors involved in information productivity and diffusion of knowledge.

To demonstrate the progress, growth and dynamics of Econophysics, the study is organized as follows: Section~\ref{sec:data} provides the data description. Section~\ref{sec:results} highlights the results which are further divided into three subsections: Subsection~\ref{subsec1} discusses the results on dynamics of citation patterns in the data and the key disciplines of the cited references. Subsection~\ref{subsec2} presents a detailed discussion on the collaboration networks at micro- (\ref{subsec2.1} and \ref{subsec2.2}), meso- (\ref{subsec2.3}) and macro- (\ref{subsec2.4}) levels. Subsection~\ref{sec:NetworkGrowth} shows the growth of co-authorship and institutional networks over years. Section~\ref{sec:summary} concludes this study and discusses the limitations and future directions.

\section{Data description}
\label{sec:data}
We collected the data from Web of Science managed (WoS) by Clarivate Analytics. The data mining API (\url{https://apps.webofknowledge.com/}) of WoS is used to fetch the records~\citep{wos}.
We searched for the papers that match the keyword \textit{Econophysics} published during 2000-2019. During 1995-1999 significant publication count is not available in WoS, so we could not perform the analysis since 1995. A total of 1458 records are retrieved including all document types. We further filtered the data based on the \textit{Document Type }and included papers which are: \textit{Articles, Reviews, Proceedings, Editorial Material, and Book Chapter} as these categories are having a sufficient number of papers. Hence, we finally filtered 1437 records. All records contain the full description of the paper like author name, affiliation, citations, publication journals, references, etc.

To retrieve the disciplines of the cited references of each paper, first, we extracted the title of each reference and then searched for that title in the WoS database. Not all cited references are listed in the WoS and this allowed us to match $74\%$ of the references.
This way we get the list of relative disciplines of all cited references in 1437 papers. To get the list of author's collaborations, we identified the author's unique ID provided by WoS (DAIS number) as there could be two authors with the same name. Similarly, corresponding to the author's ID, we identified the institutions. The corresponding author's location information is extracted from the reprinting address in the paper. Many scholars have studied the economy and the stock market behavior by using the methods of statistics, mathematics, computer science, etc. However, the focus of our study is to select papers where physics concepts have been used to study the economy and stock market behavior.

\section{Results}
\label{sec:results}
\subsection{Dynamics of citation patterns }
\label{subsec1}
\begin{figure}[!h]
   \centering
   \includegraphics[width=0.75\linewidth]{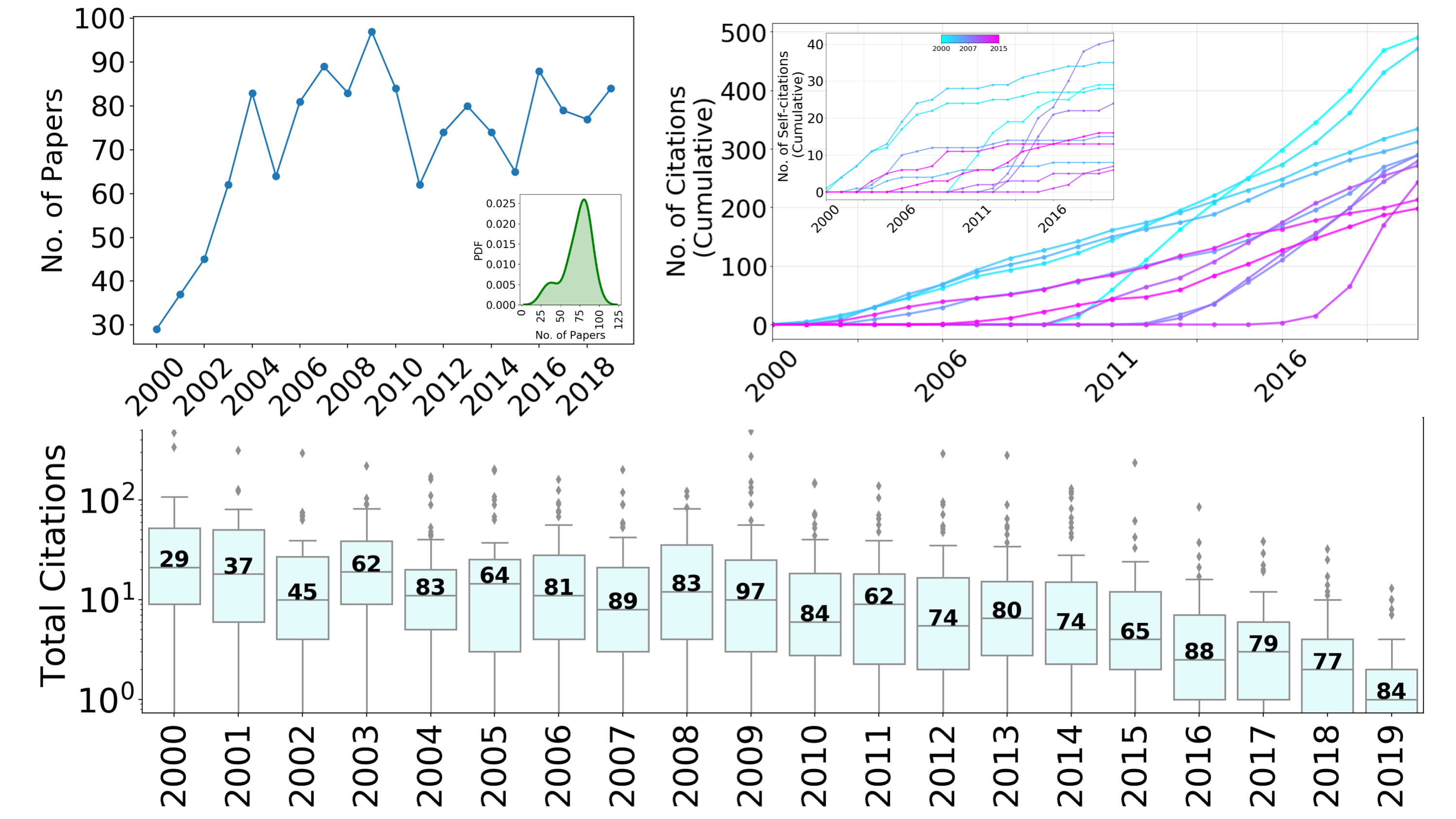}
   \llap{\parbox[b]{4.9in}{(a)\\\rule{0ex}{2.5in}}}
    \llap{\parbox[b]{0.7in}{(b)\\\rule{0ex}{2.5in}}}
    \llap{\parbox[b]{5in}{(c)\\\rule{0ex}{1.3in}}}
   \includegraphics[width=0.75\linewidth]{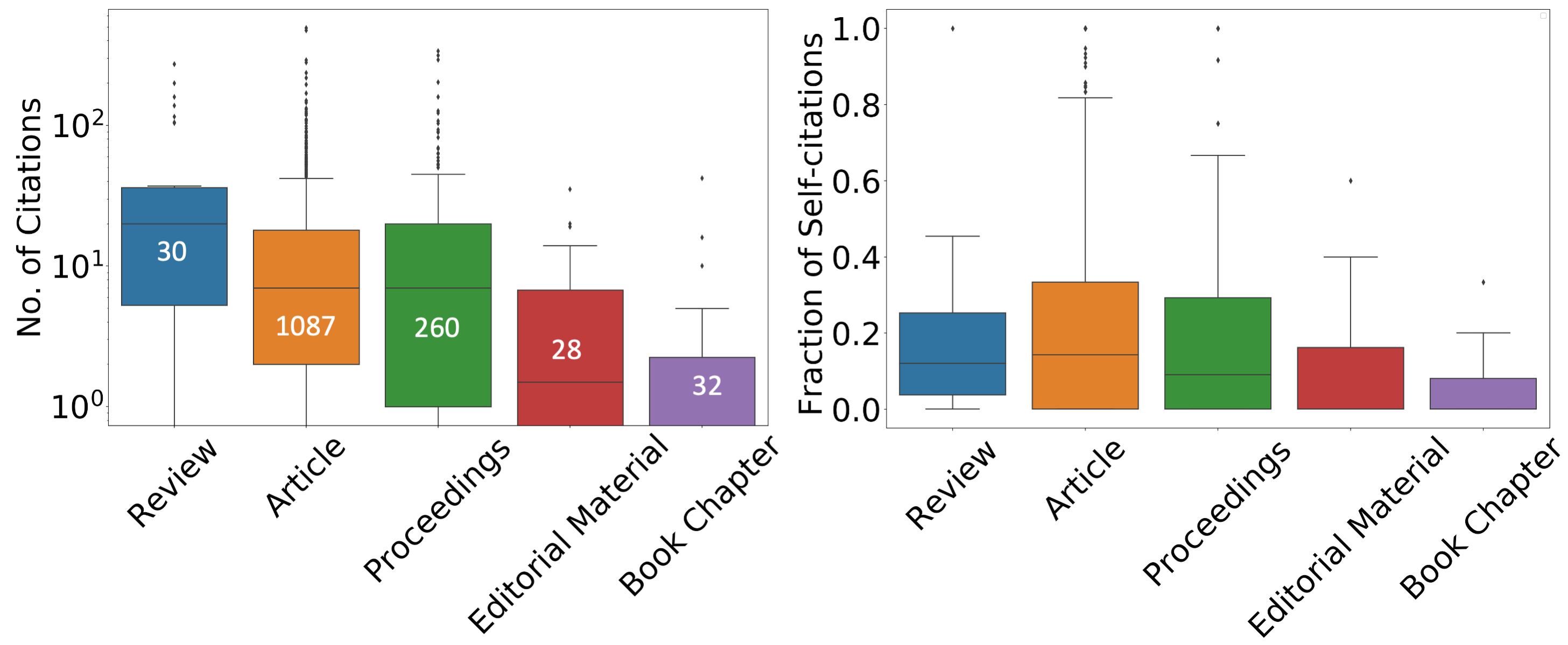}
   \llap{\parbox[b]{5in}{(d)\\\rule{0ex}{1.8in}}}
   \llap{\parbox[b]{0.5in}{(e)\\\rule{0ex}{1.7in}}}
  
   \caption{\textbf{Characterizing publications, citations, and self-citations.} (a) Total number of papers published during 2000-2019. (\textit{Inset}) Probability density of the number of papers. (b) Yearly citations growth for a few randomly selected papers published between 2000 and 2015.  \textit{(Inset)} Cumulative plot of the number of self-citations received. The color code corresponds to the publication year of each paper. (c) Citations received by papers published over years. The number inside the box shows the publication count corresponding to years. (d) The median number of citations received by different documents published during 2000-2019. The numeric value inside the box is the total number of published papers in that document category. (e) The fraction of self-citations received by each document category.
}
   \label{fig:EconoPaperCount}
\end{figure}

We have presented the characterization of number of publications, citations, self-citations, etc. in Fig.\ref{fig:EconoPaperCount}.
The number of papers published from 2000-2019 are reported in Fig.\ref{fig:EconoPaperCount}(a). To show the growth of the citations and self-citations we randomly selected a few papers published from 2000-2015. The citations received by each paper over the years since its publication is plotted cumulatively in Fig.\ref{fig:EconoPaperCount}(b). The color code depicts the publication year. The inset of the figure shows the growth of self-citations received by the same set of papers over years since its publication. Fig.\ref{fig:EconoPaperCount}(c) represents the median number of citations received by all papers published over years. The numeric value inside the box plot represents the count for the total number of papers published in the respective year. The median number of citations received by papers published as \textit{Articles}, \textit{Reviews}, \textit{Book Chapter}, etc. is shown in Fig.\ref{fig:EconoPaperCount}(d) and corresponding median self-citations is shown in Fig.\ref{fig:EconoPaperCount}(e). The numeric value inside the plot is the number of papers published. The bars are arranged according to the median number of citations rather than the number of publications. For example, papers published as \textit{Articles} and \textit{Proceedings} have received equal median number of citations; however, the number of publications as \textit{Articles} are higher than \textit{Proceedings}. On the other hand, \textit{Review}  papers are less published as compared to other document types but have received the highest median number of citations. The median self-citations received by \textit{Reviews} and \textit{Articles} are almost same.

\begin{figure}[!h]
   \centering
   \includegraphics[width=0.8\linewidth]{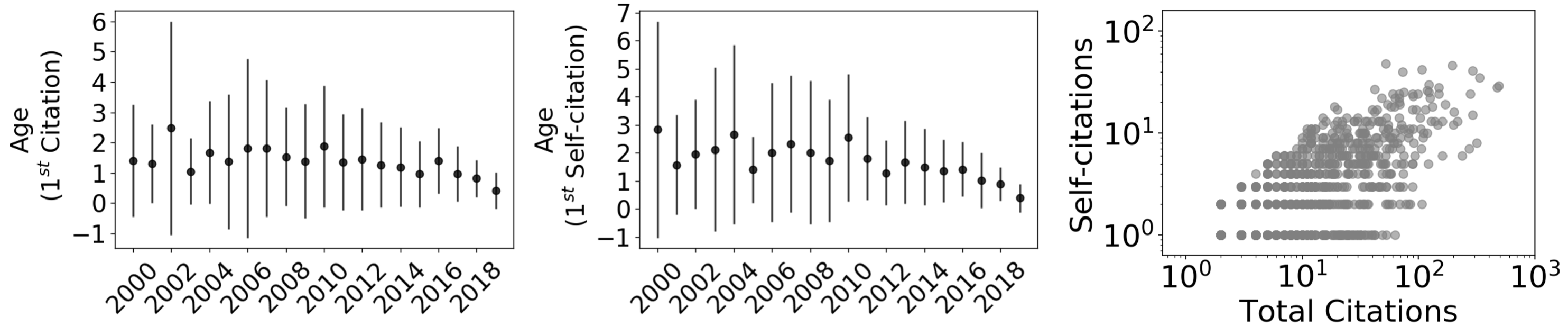}
   \llap{\parbox[b]{3.8in}{(a)\\\rule{0ex}{0.9in}}}
   \llap{\parbox[b]{2.2in}{(b)\\\rule{0ex}{0.9in}}}
   \llap{\parbox[b]{1.3in}{(c)\\\rule{0ex}{0.9in}}}
   \includegraphics[width=0.8\linewidth]{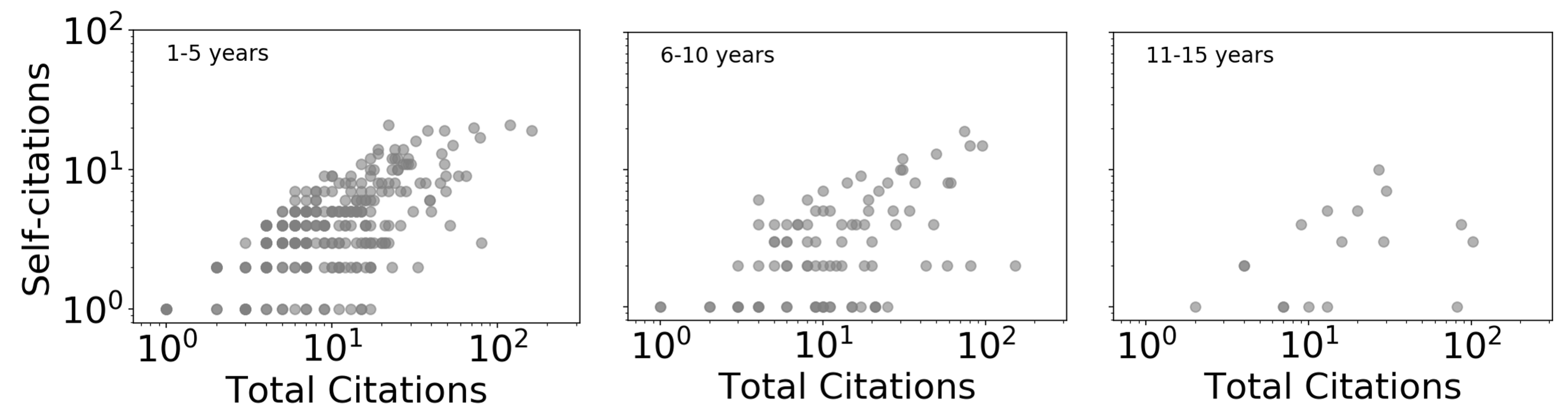}
    \llap{\parbox[b]{3.7in}{(d)\\\rule{0ex}{1.05in}}}
   \llap{\parbox[b]{2.0in}{(e)\\\rule{0ex}{1.05in}}}
	\llap{\parbox[b]{0.5in}{(f)\\\rule{0ex}{1.05in}}}

   \caption{\textbf{Dynamics of citations and self-citations over the years.} (a) Average age of the paper when it has received first citations which is not a self-citation during 2000-2019. (b) The average age of the paper when it has received the first self-citation. (c) A number of citations and self-citations received by papers from 2000-2019. (d) Relationship between the citations and self-citations received by each paper in the first five years after publication. (e-f) Five-five years' time interval behavioral change in citations and corresponding self-citations.}
\label{fig:Fig2}
\end{figure}

Fig.\ref{fig:Fig2} represents the dynamics of citations and self-citations over the years. 
Fig.\ref{fig:Fig2}(a) shows the average age of a paper when it has received the first citation which is not a self-citation during 2000-2019. Similarly, the average age of a paper when it has received first self-citation is shown in Fig.\ref{fig:Fig2}(b). On an average, the paper receives first citation and self-citation within the first two years after its publication. Fig.\ref{fig:Fig2}(c) shows the overall citations and self-citations received by papers from 2000-2019. Higher the number of citations, the higher the self-citations. During the first five years of a publication, the count of self-citations has increased with the increase of citations as shown in Fig.\ref{fig:Fig2}(d)~\citep{fowler2007does}. This shows that during the initial year's authors tend to cite their papers quite often to maintain the visibility of the papers. This association decreases with the increase of the time interval (Fig.\ref{fig:Fig2}(e-f)).

\subsubsection{Referencing disciplines in the papers}
\label{subsec1.1}

\begin{figure}[!h]
   \centering
        \includegraphics[width=0.44\linewidth] {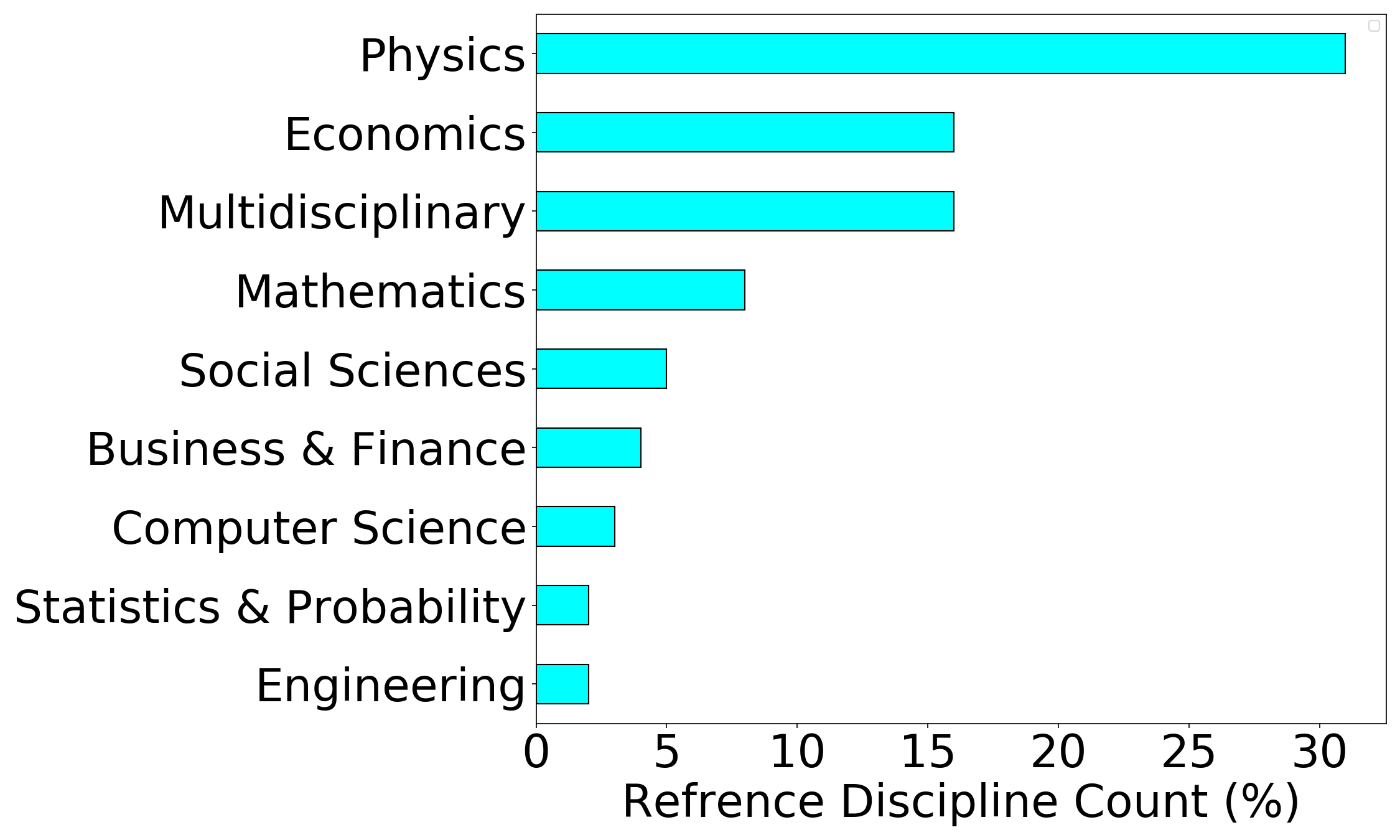}
     \llap{\parbox[b]{2.8in}{(a)\\\rule{0ex}{1.6in}}}
    \includegraphics[width=0.48\linewidth] {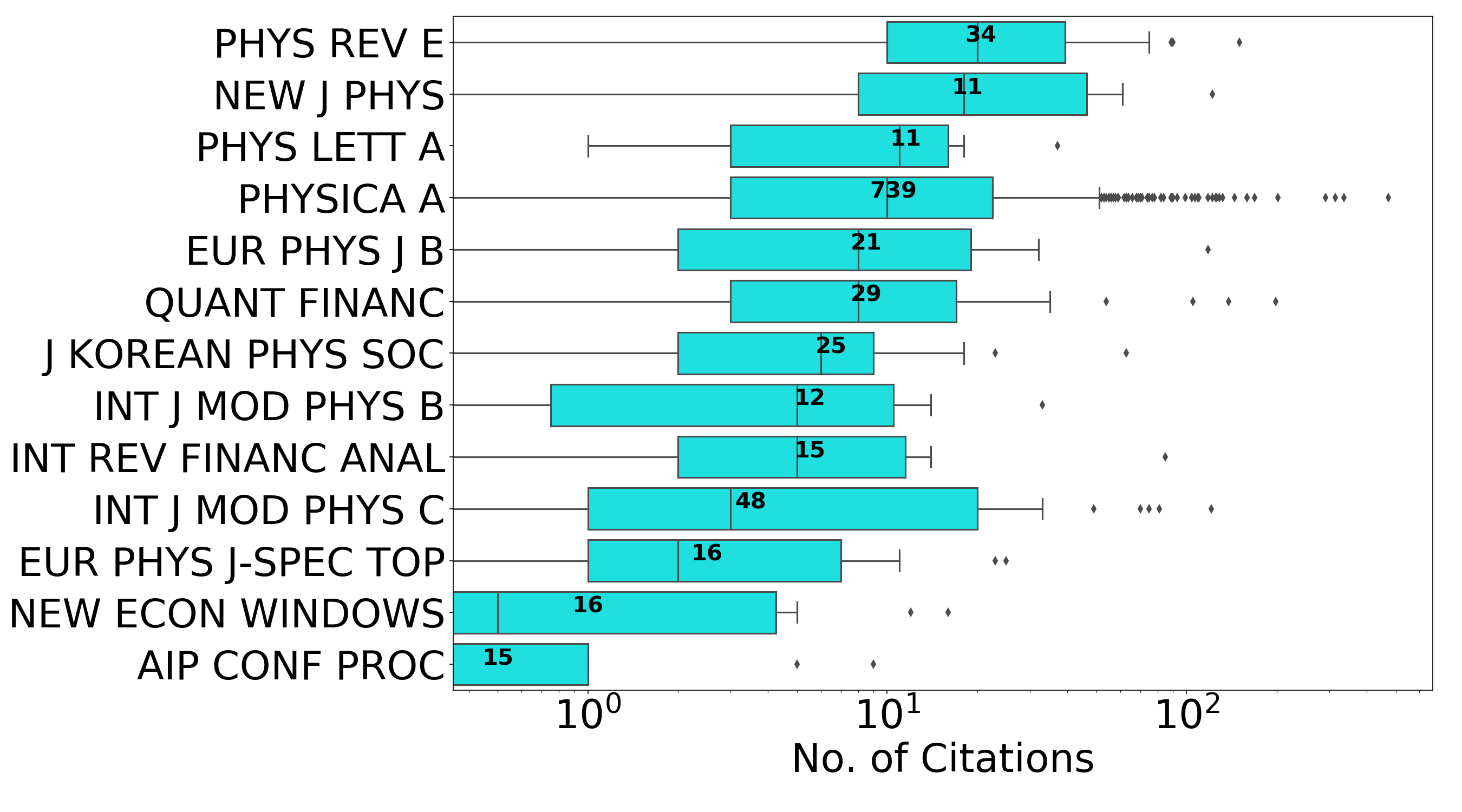}
     \llap{\parbox[b]{3in}{(b)\\\rule{0ex}{1.6in}}}

   \caption{\textbf{Key disciplines by cited references and publication journals.} (a) Bar plot shows the number of times (in $\%$) a reference has been cited from a discipline. $31\%$ of the references are cited from physics discipline and $16\%$ are from economics, which clearly depicts the true nature of Econophysics. (b) The median number of citations received by different journals that published Econophysics papers. The numeric value inside the box is the total number of published papers. The bars are arranged according to the median of citations rather than the number of citations. The majority of the papers are published in physics journals.}
\label{fig:Fig3}
\end{figure}

To understand which disciplines have contributed more to the growth of Econophysics, we analyzed the references cited by each paper. We retrieved the disciplines of all the cited references and analyzed the contribution of disciplines. Fig.\ref{fig:Fig3}(a) highlights the disciplines according to the number of cited references (in $\%$). It is evident that major references were quoted from \textit{Physics} followed by \textit{Economics} which clearly represents the true nature of Econophysics. The proportion of physics references also revealed the major contribution of physicists' in the field. Fig.\ref{fig:Fig3}(b) highlights the journals based on the median number of citations received by the papers. The bars are arranged according to the median of citations rather than the number of citations. \textit{Physica A} has published more papers (739) than \textit{Physics Review E } (34); however, \textit{Physics Review E } has received higher the median number of citations (20) than \textit{Physica A} (10). The first few journals are also physics-based journals where papers have gained higher citations. 

\subsection{Collaboration network}
\label{subsec2}
Here we presented the scientific collaborations at micro, meso, and macro-level.

\subsubsection{Micro-level analysis: Author's collaboration network}
\label{subsec2.1}
\begin{figure}[!h]
   \centering
   \includegraphics[width=0.75\linewidth]{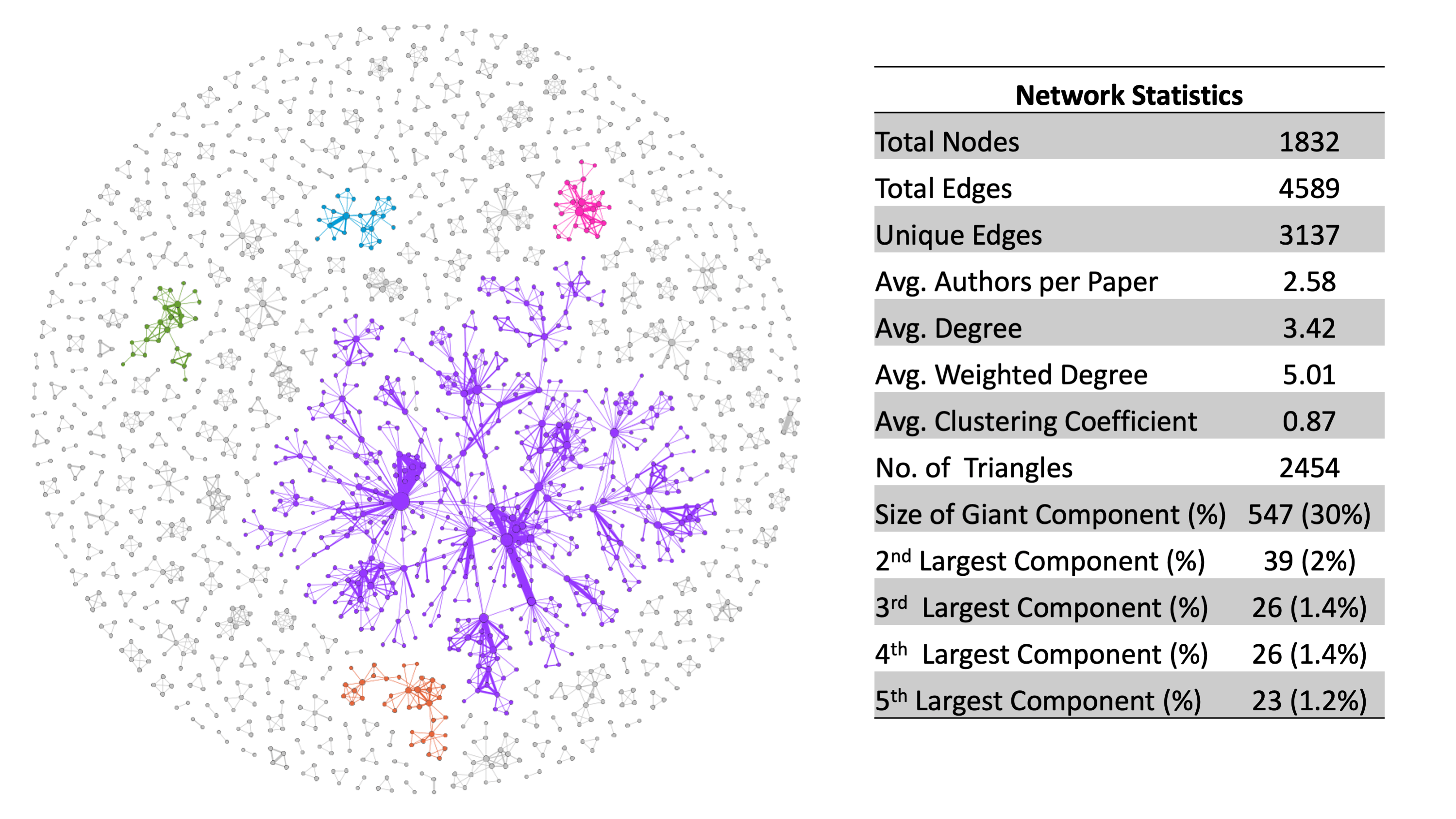}
	  \llap{\parbox[b]{5.0in}{(a)\\\rule{0ex}{2.1in}}}
   \includegraphics[width=0.75\linewidth]{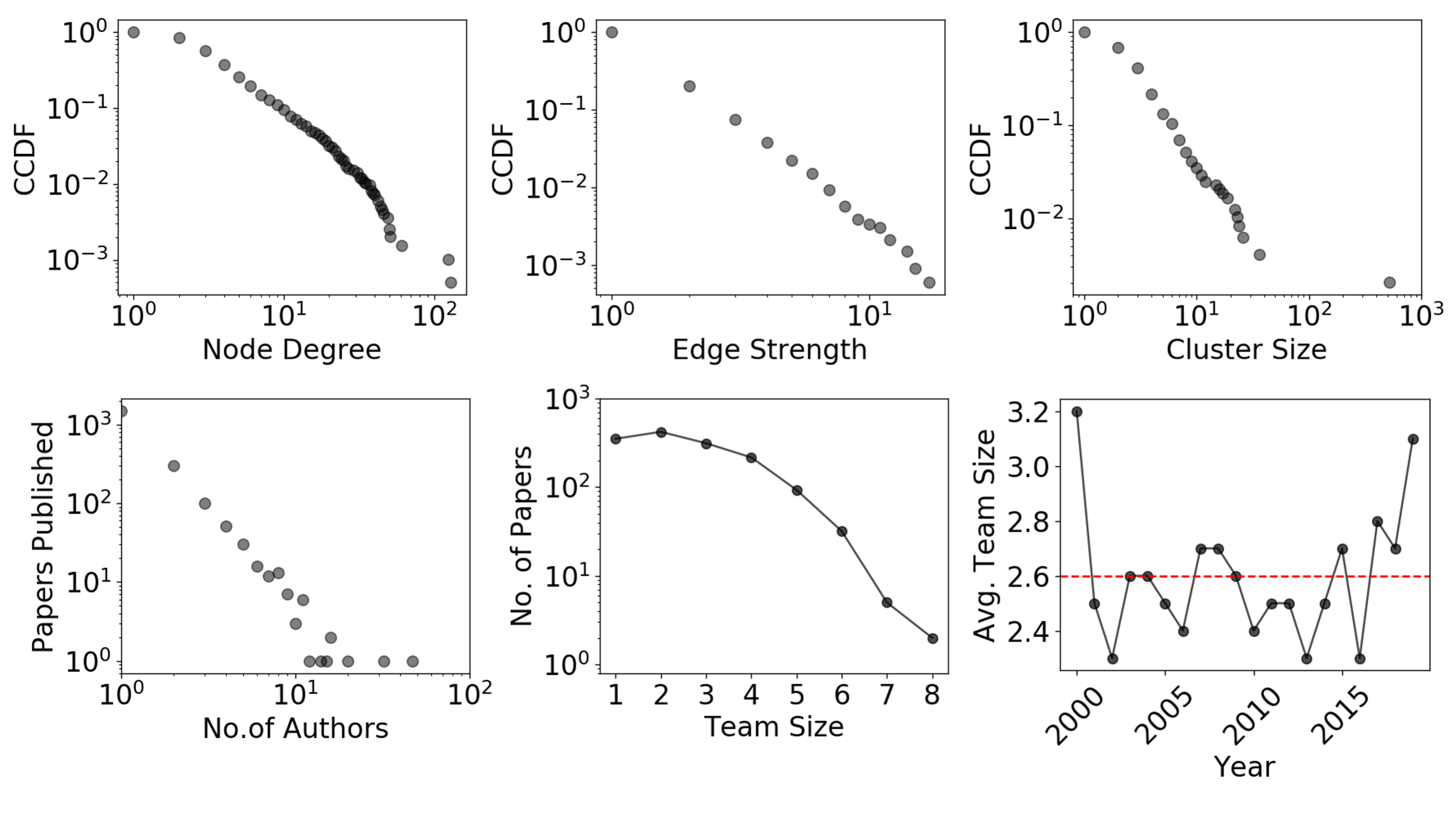}
   \llap{\parbox[b]{3.6in}{(b)\\\rule{0ex}{2.4in}}}
   \llap{\parbox[b]{2.0in}{(c)\\\rule{0ex}{2.4in}}}
   \llap{\parbox[b]{0.55in}{(d)\\\rule{0ex}{2.4in}}}
   \llap{\parbox[b]{3.7in}{(e)\\\rule{0ex}{1.15in}}}
   \llap{\parbox[b]{2.2in}{(f)\\\rule{0ex}{1.15in}}}
   \llap{\parbox[b]{0.6in}{(g)\\\rule{0ex}{1.15in}}}
   
    \caption{\textbf{Co-authorship network.} (a) An undirected weighted co-authorship network having 1834 nodes and 4590 edges (3137 unique edges). The nodes represent authors and edges represent the collaboration among authors. We have filtered the self-loops in the network representation. The size of the node corresponds to the weighted degree of the node and the width of the edge represents the strength of the collaboration.  Different colors represent the first five largest connected components. The giant component (colored in purple) contains 547 nodes which are $30\%$ of the total nodes of the network. (b-d) The statistical properties of the network as complementary cumulative density functions (CCDF's): weighted degree, edge weight, and cluster size, respectively.
(e) Number of papers published by authors represents the contribution of authors in the field. A few authors have published a large number of papers. (f) Papers published by teams of varying sizes. (g) Time evolution of the typical number of team members. The red line represents the average team size. The table shows the network statistics. The network is constructed in \textit{Gephi 0.9.2}.
}
\label{fig:Fig4}
\end{figure}

In the co-authorship network~\citep{fan2004network}, we have constructed an undirected weighted network consists of 1834 nodes and 4590 edges (3137 unique edges) as shown in Fig.\ref{fig:Fig4}(a), where nodes correspond to authors and edges represent the collaboration (when two or more authors write a paper together). Single-authored articles are excluded from the data set since they do not contribute to the co-authorship network. The first five largest connected components of the network are colored differently. The giant component (colored in purple) contains the $30\%$ of the total nodes of the network. The giant component is further elaborated in Fig.\ref{fig:Fig5}. The second-largest component (colored in green) contains $2\%$ nodes, and so on (see network statistics table in Fig.\ref{fig:Fig4}). It is often perceived that certain authors are actively engaged in collaboration than others.
Fig.\ref{fig:Fig4}(b-c) shows the complementary cumulative density function (CCDF) of the degree of the nodes and edges strength which represents the author's collaborations and the strength of the collaboration. The power-law behavior of CCDF shows that there are few authors who share a large number of collaborations.
The CCDF's of the cluster size or connected components are shown in Fig.\ref{fig:Fig4}(d). The power-law behavior of the cluster size distribution clearly shows that only one component contains a large number of nodes. In the network, $55\%$ of nodes have clustering coefficient 1, and $28\%$ have 0. The highest clustering coefficient represents how well the nodes are connected to their neighbors. The highest average clustering coefficient (0.87) shows that almost everyone is connected to others in the network. Fig.\ref{fig:Fig4}(e) shows the relation between the number of authors and the number of papers published by them. A few authors have published more than 10 papers, whereas a large number of authors have published less than 10 papers. Also, how big is the team size of authors is studied in Fig.\ref{fig:Fig4}(f). The majority of the papers are either published as a single author or two authors. There are few papers that have been written by 7 to 8 authors which is also the largest team size. Fig.\ref{fig:Fig4}(g) shows the evolution of team size in scientific collaborations~\citep{guimera2005team}. Over the years the team size fluctuates from an average of 2 to an average of 3.

\subsubsection{Giant component }
\label{subsec2.2}

\begin{figure}[!h]
   \centering
   \includegraphics[width=0.75\linewidth]{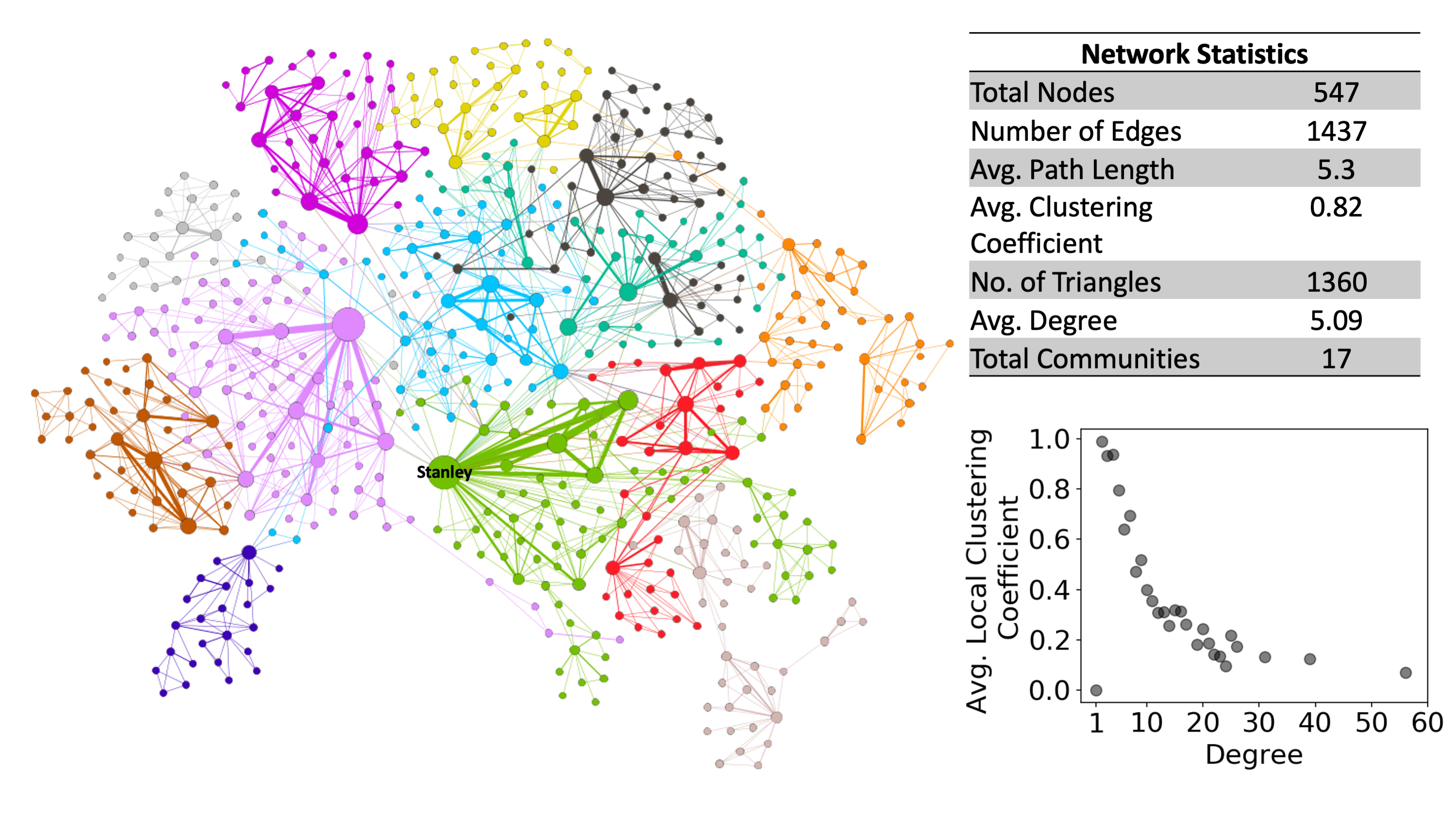}
  \llap{\parbox[b]{5.1in}{(a)\\\rule{0ex}{2.0in}}}
     \llap{\parbox[b]{1.9in}{(b)\\\rule{0ex}{1.2in}}}
    \includegraphics[width=0.75\linewidth]{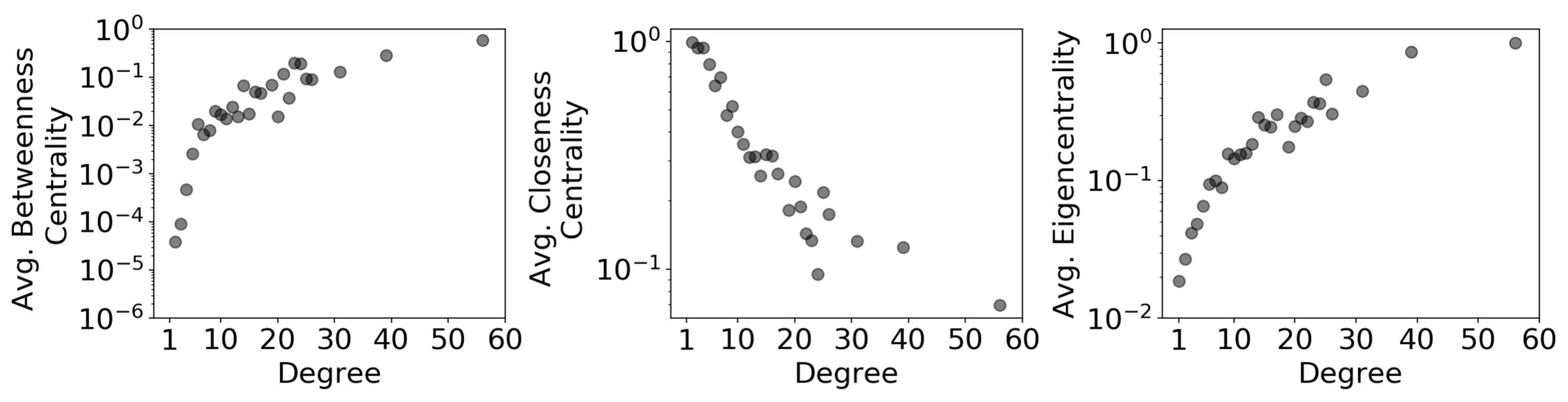}
	\llap{\parbox[b]{4.8in}{(c)\\\rule{0ex}{1.2in}}}
   \llap{\parbox[b]{3.2in}{(d)\\\rule{0ex}{1.2in}}}
   \llap{\parbox[b]{1.6in}{(e)\\\rule{0ex}{1.2in}}}

    \caption{\textbf{A giant component of the co-authorship network.}  (a) A zoomed-in view of the giant component. A modularity detection algorithm has been used to detect the communities among the network. The node with the largest connectivity is labeled by the author's name. (b) Degree versus the average local clustering coefficients of the nodes. On an average nodes of higher degrees exhibit lower local clustering. (c) Degree versus average betweenness centrality of the nodes. Nodes with higher betweenness centrality represent the potential key authors and nodes with higher degree represent hubs in the network. It highlights that the nodes with the highest degree act as the bridge to compute the shortest-path among all nodes in the network. (d) Degree versus average closeness centrality of the nodes. On an average, nodes of higher degrees share a low closeness. (e) Degree versus average eigencentrality of the nodes. The eigencentrality measures the prestige of the node in the network. On an average, nodes of higher degrees have higher prestige. The table shows the network statistics. The network is constructed in \textit{Gephi 0.9.2}.
}
\label{fig:Fig5}
\end{figure}

Fig.\ref{fig:Fig5}(a) shows the zoomed-in view of the giant component of the co-authorship network extracted from Fig.\ref{fig:Fig4}(a). A modularity maximization algorithm is used to find out the communities inside the giant component~\citep{chen2014community}. Different colors represent different communities in the giant network. There is a total of 17 communities in the network and the node with the highest number of connections is labeled with the author's name. Also, the average path length of the network is 5.3 which reveals the ``small-world'' property of the network~\citep{watts1998collective}. In the community of econophysicists', everyone is connected to others in $ \approx$ 5 steps.
The relationship between nodes degree and local clustering coefficient is shown in Fig.\ref{fig:Fig5}(b)). On an average, nodes of higher degree exhibit lower local clustering. A list of top 50 authors based on the degree (collaboration) is shown in Table\ref{table:50Aauthors}.
Fig.\ref{fig:Fig5}(c-e) represents the relationship among nodes degree and the centrality measures: betweenness, closeness, and eigencentrality. On an average, nodes of higher degree exhibit higher betweenness and eigencentrality but lower closeness centrality. Nodes with higher betweenness centrality represent the potential key authors in the network, whereas nodes with higher degree represent hubs in the network. It highlights that the nodes with the highest degree act as the bridge to compute the shortest-path among all nodes in the network. Also, the largest value of eigencentrality represents the prestige/influence of the node in the network.

\subsubsection{Meso-level analysis: Authors' affiliations network}
\label{subsec2.3}

\begin{figure}[!h]
   \centering
   \includegraphics[width=0.75\linewidth]{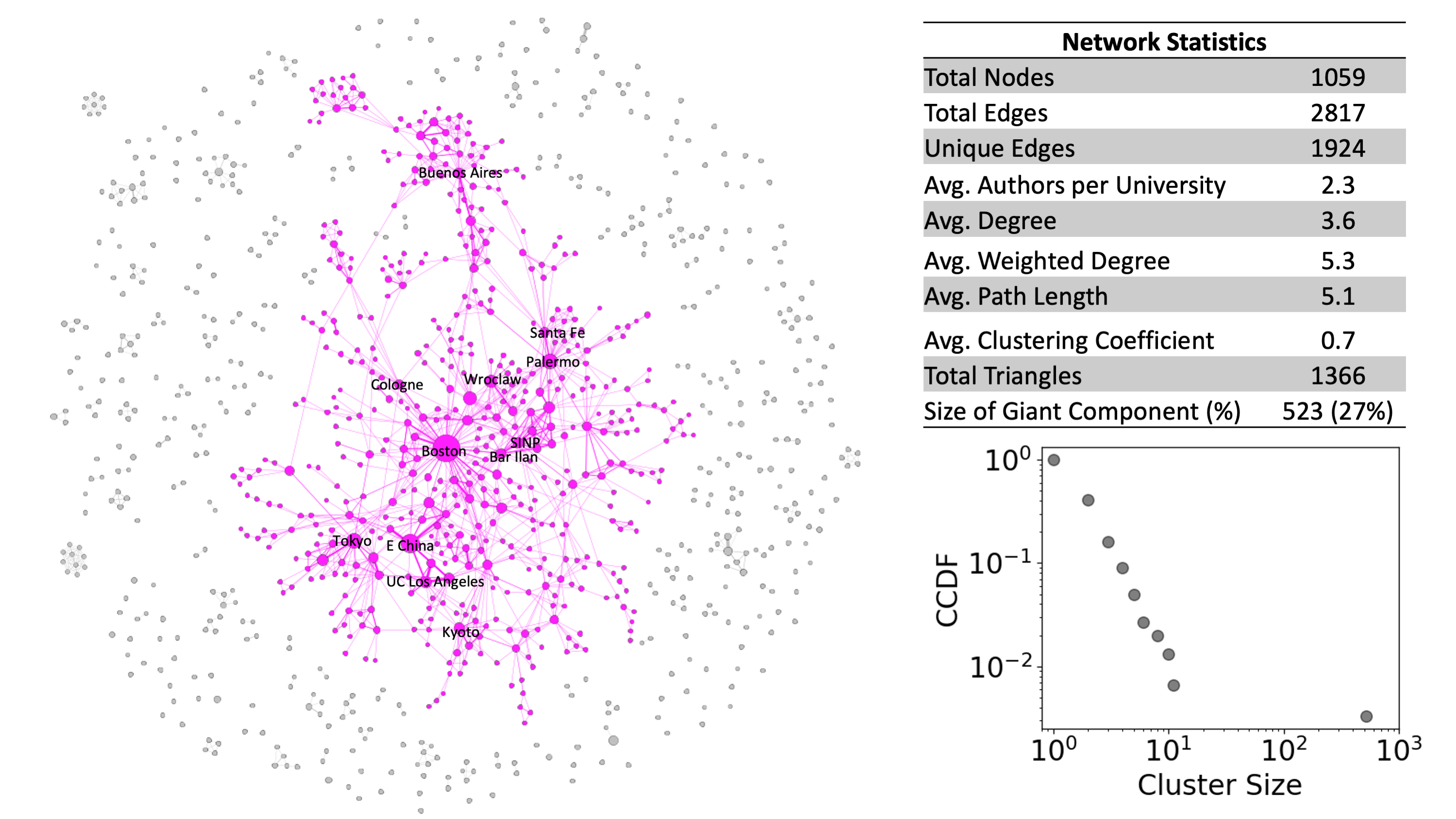}
    \llap{\parbox[b]{5.1in}{(a)\\\rule{0ex}{2.0in}}}
     \llap{\parbox[b]{0.6in}{(b)\\\rule{0ex}{1in}}}
    \includegraphics[width=0.75\linewidth]{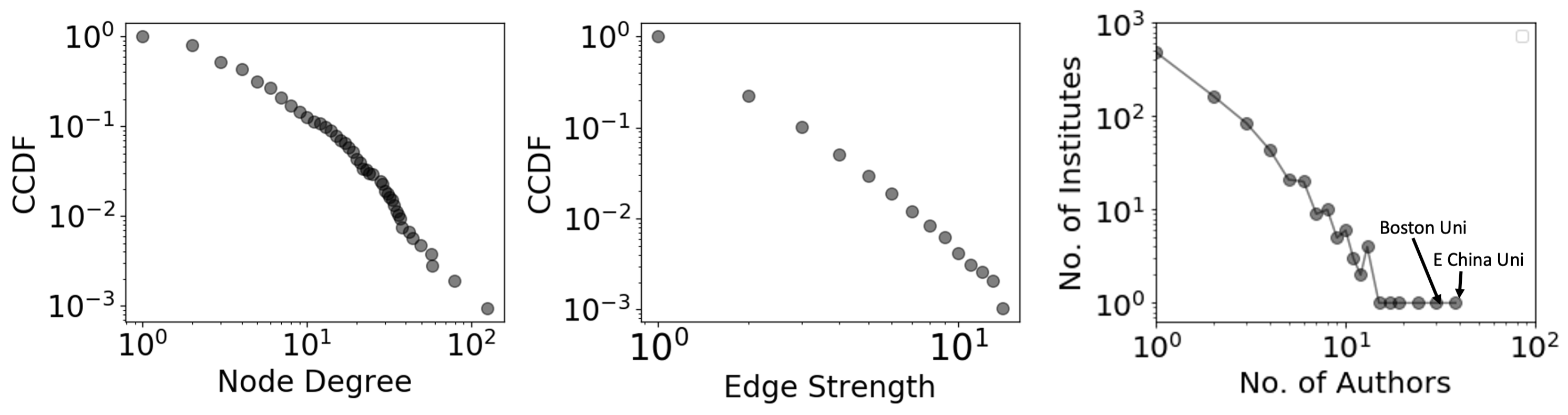}
\llap{\parbox[b]{3.6in}{(c)\\\rule{0ex}{1.0in}}}
   \llap{\parbox[b]{2.0in}{(d)\\\rule{0ex}{1.0in}}}
   \llap{\parbox[b]{0.5in}{(e)\\\rule{0ex}{1.0in}}}

   \caption{\textbf{Author's affiliations network.} (a) An undirected weighted network of institutions having 1059 nodes and 2817 edges (1924 unique edges) where nodes represent the institutions and edges represent the collaboration among the institutes across the globe. The giant component (colored in dark pink) comprises $27\%$ of the nodes. The institutes with strong collaborations are labeled with the names. There is an isolated institution in the network that corresponds to within institution collaboration; however, we have filtered the self-loops in the network representation. The size of the node represents the weighted degree and the width of the edge represents the collaboration strength. (b) CCDF of cluster size. (c) CCDF of nodes degree. (d) CCDF of edges strength. (e) A number of authors corresponding to a number of institutions. A large number of authors correspond to a few institutions. The table shows the network statistics. The network is constructed in \textit{Gephi 0.9.2}.
}
\label{fig:Fig6}
\end{figure}

After the institutionalization of Econophysics in 1995, many reputed institutes have initialized research on it and some institutes have started courses on it~\citep{dash2015judging, ortega2013institutional}. To investigate the contribution of different institutions, an undirected weighted authors' affiliations (institutions) network is constructed (see Fig.\ref{fig:Fig6} (a)). The network consists of 1059 institutions/universities and shows 2817 possible collaborations between institutions across the globe. Self-loops are removed while plotting the network, however, included in the analysis. The giant component (colored in dark pink) contains $27\%$ of the network nodes shown in Fig.\ref{fig:Fig6}(b) as CCDF of cluster size. Fig.\ref{fig:Fig6}(c-d)) show the CCDF's of nodes degree and edges strength, respectively. Fig.\ref{fig:Fig6} (e) shows the number of authors corresponding to the number of institutions working on Econophysics. A large number of authors belong to a few institutions. The top two institutions in terms of the number of authors and collaborations are \textit{East China University of Science and Technology} (ECUST) and \textit{Boston University}. \textit{ECUST} produces a large number of authors, whereas Boston University shares a large number of collaborations (see Table~\ref{table:50Uni} for institutions details).

\subsubsection{Macro-level analysis: Countries' collaborations network}
\label{subsec2.4}
\begin{figure}[!h]
   \centering
      \includegraphics[width=0.85\linewidth] {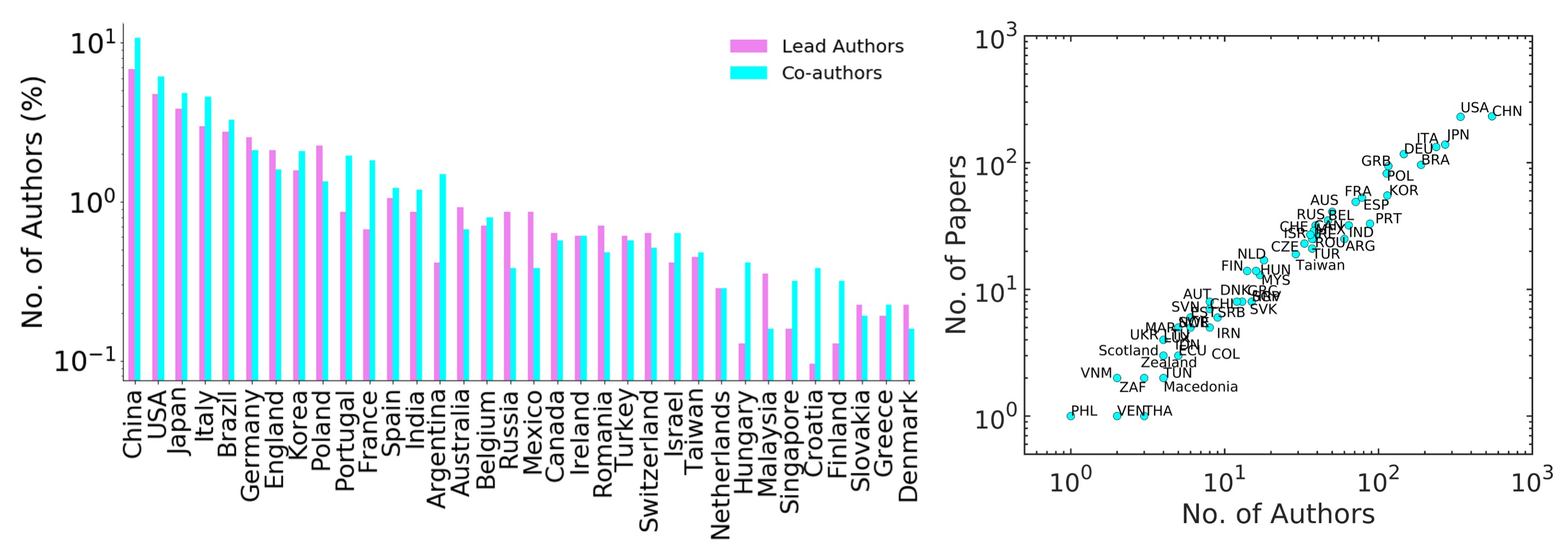}
       \llap{\parbox[b]{5.6in}{(a)\\\rule{0ex}{1.8in}}}
       \llap{\parbox[b]{2.45in}{(b)\\\rule{0ex}{1.8in}}}
      \includegraphics[width=0.75\linewidth] {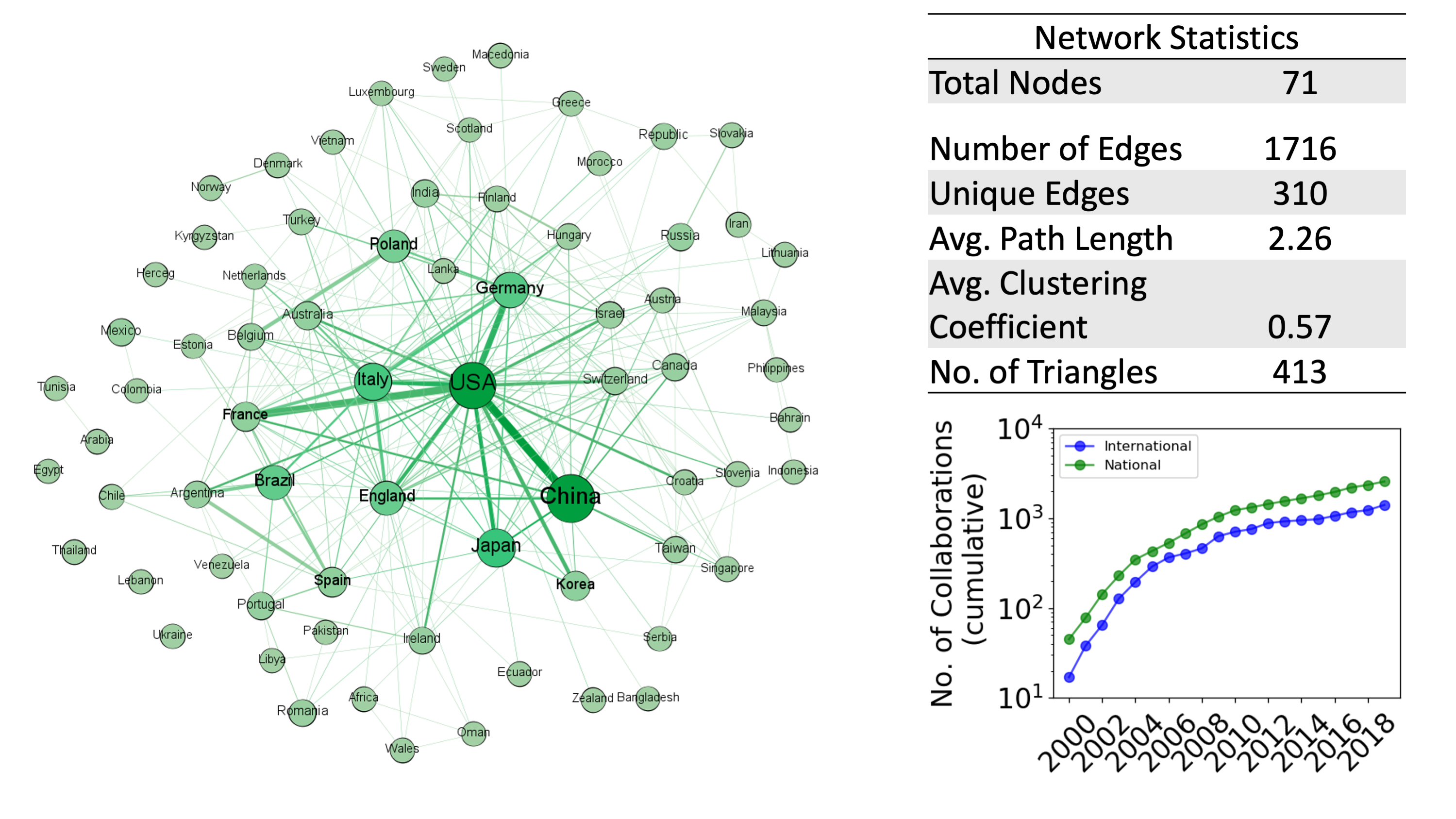}
       \llap{\parbox[b]{4.5in}{(c)\\\rule{0ex}{2.2in}}}
       \llap{\parbox[b]{2.1in}{(d)\\\rule{0ex}{1.2in}}}

   \caption{\textbf{countries collaborations network.} (a) Number of papers published as a corresponding author (colored violet) and as one of the authors (colored cyan) listed for few countries. The countries are arranged in descending order based on the number of corresponding authors. (b) Scattered plot for 71 countries representing the number of papers published by authors. (c) An undirected weighted network of countries corresponding to the author's location contains 71 nodes and 1716 edges (310 unique edges) where nodes represent countries and edges represent the scientific collaboration. There are a few isolated countries too. For simplicity, we filtered the self-loops from the network representation which correspond to within the country collaboration. The size of the node represents the weighted degree and the color gradient of the nodes varies according to the degree. The edge width represents the number of connections/collaborations among the nodes. The cross-country network shows the countries that have strong ties among them. (d) The evolution of the cumulative growth of international and national collaborations. The table shows the network statistics. The network is constructed in \textit{Gephi 0.9.2}.}
   \label{fig:Fig7}
\end{figure}

To visualize the expansion of the econophysicists' across the globe we have studied the geolocations of authors. Fig.\ref{fig:Fig7}(a) represents the number of authors in different countries' (in $\%$) working on Econophysics. The violet-colored bars represent the corresponding authors who lead the projects and cyan colored bars represent the co-authors of the papers. Here, we displayed results only for few countries' as per the number of corresponding authors. China is leading in terms of the number of corresponding as well as co-author's participation.
Fig.\ref{fig:Fig7}(b) highlights the number of papers published by the number of authors in the respective countries. The results are presented in 71 countries. The trend reveals the signature of scaling behavior in terms of the author's publications across the globe.
Further, an undirected weighted network of countries' with 71 nodes and 1716 edges is constructed in Fig.\ref{fig:Fig7}(c). There are self-loops present in the network which correspond to either a single author paper or collaboration among the same country.  The size of the node represents the number of authors in the respective country and the edge width represents the number of times a collaboration occurred. Results highlight a strong collaboration between the USA and France; however, the number of authors is higher in the USA rather than in France which shows there might be a small but active community of researchers in the field. We also find that the within-country collaboration is more active as compared to cross-country. Hence, China, the USA, Italy, Japan, Germany, France, etc. have a large number of authors, a large number of publications, and strong connectivity/ collaboration among them (self-loops not shown in-network). We can say that these leading countries' are driving the discipline, however, other countries are also contributing to the growth of the discipline and getting connecting to the leading countries. Fig.\ref{fig:Fig7}(d) shows the evolution of the cumulative growth of international and national collaborations. Results highlights that national collaboration is higher than international collaborations. China shares more nationals, whereas the USA shares more international collaborations. There is a dip in the international collaborations trend during 2007-2008, this was the time when the stock market crashed due to the bankruptcy of Lehman Brothers.

\subsection{Network growth}
\label{sec:NetworkGrowth}
\begin{figure}[!h]
   \centering
     \includegraphics[width=0.55\linewidth] {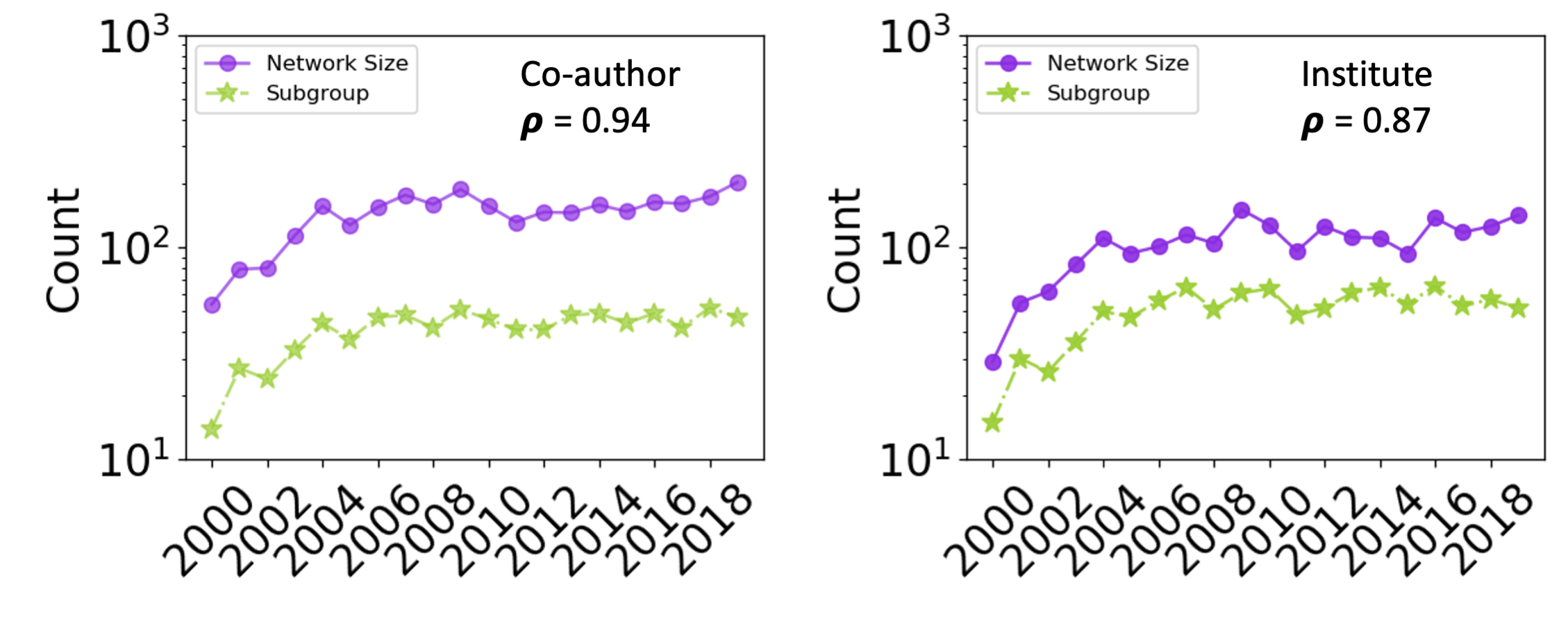}
      \llap{\parbox[b]{3.8in}{(a)\\\rule{0ex}{1.3in}}}
   \llap{\parbox[b]{1.85in}{(b)\\\rule{0ex}{1.3in}}}
      \includegraphics[width=0.75\linewidth] {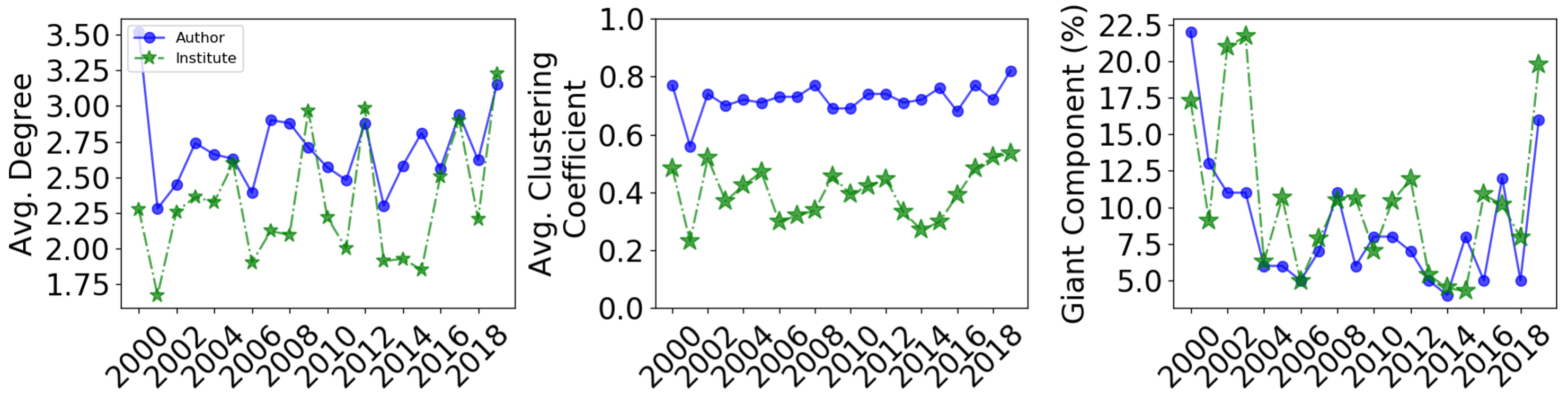}
\llap{\parbox[b]{3.6in}{(c)\\\rule{0ex}{1in}}}
   \llap{\parbox[b]{2in}{(d)\\\rule{0ex}{1.05in}}}
   \llap{\parbox[b]{0.5in}{(e)\\\rule{0ex}{1.0in}}}
 
   \caption{\textbf{Network growth over the years.} (a-b) Size of the network and number of connected components of co-authorship and institutional networks over years, respectively. The time series of network size evolution is highly correlated with the time series of the evolution of the number of connected components for both the networks. (c-e) Growth of co-authorship network and affiliation network over years: (c) average degree, (d) average clustering coefficient, and (e) size of the giant component (in \%).}
   \label{fig:Fig8}
\end{figure}

The evolution of fundamental statistical properties of the scientific collaboration networks in terms of the average degree of the nodes ($<k>$), average clustering coefficient ($<cc>$), and size of the giant component (GC(\%)) during 2000-2019 is shown in Fig.\ref{fig:Fig8}. The evolution of the co-authorship network is shown in Fig.\ref{fig:Fig8}(a) where the time series of network growth and a number of connected components shares a high amount of correlation (0.94). Similarly, the evolution of the author's affiliation network is shown in Fig.\ref{fig:Fig8}(b) where the time series of network growth and a number of connected components also shares a high amount of correlation (0.87).
The evolution of the network's average degree, average clustering coefficient, and size of the giant component (in\%) are shown for both the co-authorship and affiliations networks in Fig.\ref{fig:Fig8}(c-e), respectively. On an average, the degree of the co-authorship network varies between 2 to 3, and the clustering coefficient varies between 0.6 to 0.8. Similarly, on an average, the degree of the affiliations network varies between 1 to 3 and the clustering coefficient varies between 0.2 to 0.6 over years. The average path length of the network lies between 2 to 3 which reveals the ``small-world'' behavior of the network at every time step. A higher average clustering coefficient shows that nodes are grouped into communities.

\begin{table}[!h]
\caption{List of 50 authors based on the degree (collaboration).
The table shows the Author name. country, affiliation, number of collaboration ($k$), and clustering coefficient ($cc$).}
\resizebox{\columnwidth}{!}{%
\begin{tabular}{|l|l|l|l|l|l|l|l|l|l|l|l|}
\hline
\textbf{\begin{tabular}[c]{@{}l@{}}S. \\ No.\end{tabular}} & Country                          & Author        & Affiliation                                                                                  & $k$ & $cc$    & \textbf{\begin{tabular}[c]{@{}l@{}}S. \\ No.\end{tabular}} & Country     & Author          & Affiliation                                                                                     & $k$  & $cc$    \\ \hline
\textbf{1}                                                 & USA                              & Stanley, HE   & Boston University                                                                        & 56 & 0.07 & \textbf{26}                                                & China       & Gu, GF          & \begin{tabular}[c]{@{}l@{}}East China University of\\ Science \& Technology\end{tabular}    & 15 & 0.45 \\ \hline
\textbf{2}                                                 & China                            & Zhou, WX      & Chinese Academy of Sciences                                                              & 39 & 0.13 & \textbf{27}                                                & South Korea & Jung, WS        & \begin{tabular}[c]{@{}l@{}}Pohang University of \\ Science \& Technology\end{tabular}       & 15 & 0.34 \\ \hline
\textbf{3}                                                 & Japan                            & Takayasu, H   & \begin{tabular}[c]{@{}l@{}}Sony Computer   \\ Science Laboratories\end{tabular}          & 31 & 0.13 & \textbf{28}                                                & USA         & Johnson, NF     & University of Miami                                                                         & 15 & 0.16 \\ \hline
\textbf{4}                                                 & Italy                            & Mantegna, RN  & University of Palermo                                                                    & 26 & 0.17 & \textbf{29}                                                & England     & Schinckus, C    & University Leicester Finance                                                                & 14 & 0.26 \\ \hline
\textbf{5}                                                 & China                            & Ren, F        & \begin{tabular}[c]{@{}l@{}}E China University of  \\ Science \& Technology\end{tabular}  & 25 & 0.22 & \textbf{30}                                                & China       & Jiang, XF       & \begin{tabular}[c]{@{}l@{}}Collaborat Innovat \\ Ctr Adv Microstruct\end{tabular}           & 13 & 0.32 \\ \hline
\textbf{6}                                                 & Belgium                          & Ausloos, M    & University of Liege                                                                      & 24 & 0.10 & \textbf{31}                                                & China       & Zheng, B        & \begin{tabular}[c]{@{}l@{}}Collaborat Innovat \\ Ctr Adv Microstruct\end{tabular}           & 13 & 0.26 \\ \hline
\textbf{7}                                                 & England                          & Di Matteo, T  & Kings College London                                                                     & 23 & 0.13 & \textbf{32}                                                & China       & Zhang, W        & Tianjin University                                                                          & 13 & 0.59 \\ \hline
\textbf{8}                                                 & \multicolumn{1}{c|}{Switzerland} & Sornette, D   & Swiss Finance Institute                                                                  & 22 & 0.10 & \textbf{33}                                                & Croatia     & Podobnik, B     & University of  Rijeka                                                                       & 13 & 0.24 \\ \hline
\textbf{9}                                                 & Japan                            & Takayasu, M   & Tokyo Institute of Technology                                                            & 22 & 0.19 & \textbf{34}                                                & England     & Preis, T        & University College London                                                                   & 13 & 0.33 \\ \hline
\textbf{10}                                                & Japan                            & Kaizoji, T    & Int Christian University                                                                 & 21 & 0.20 & \textbf{35}                                                & India       & Chakrabarti, BK & Saha Institute of Nuclear Physics                                                           & 13 & 0.24 \\ \hline
\textbf{11}                                                & USA                              & Yakovenko, VM & University of Maryland                                                                   & 21 & 0.18 & \textbf{36}                                                & Ireland     & McCauley, JL    & NUI Galway                                                                                  & 13 & 0.26 \\ \hline
\textbf{12}                                                & China                            & Qiu, T        & Nanchang Hangkong University                                                             & 20 & 0.24 & \textbf{37}                                                & Japan       & Mizuno, T       & University of Tsukuba                                                                       & 13 & 0.33 \\ \hline
\textbf{13}                                                & Ireland                          & Richmond, P   & Univ Dublin Trinity College                                                              & 19 & 0.17 & \textbf{38}                                                & South Korea & Kim, SY         & \begin{tabular}[c]{@{}l@{}}Korea Advance Institute of \\ Science \& Technology\end{tabular} & 13 & 0.24 \\ \hline
\textbf{14}                                                & Italy                            & Gallegati, M  & Univ Politecn Marche                                                                     & 19 & 0.22 & \textbf{39}                                                & Australia   & Aste, T         & Australian National University                                                              & 12 & 0.21 \\ \hline
\textbf{15}                                                & South Korea                      & Lee, JW       & Inha University                                                                          & 19 & 0.15 & \textbf{40}                                                & China       & Wang, GJ        & Hunan University                                                                            & 12 & 0.26 \\ \hline
\textbf{16}                                                & Canada                           & Li, SP        & University of Toronto                                                                    & 17 & 0.20 & \textbf{41}                                                & China       & Xie, C          & Hunan University                                                                            & 12 & 0.30 \\ \hline
\textbf{17}                                                & China                            & Jiang, ZQ     & \begin{tabular}[c]{@{}l@{}}East China University of\\ Science \& Technology\end{tabular} & 17 & 0.33 & \textbf{42}                                                & South Korea & Yang, JS        & Korea University                                                                            & 12 & 0.36 \\ \hline
\textbf{18}                                                & China                            & Huang, JP     & Fudan University                                                                         & 17 & 0.23 & \textbf{43}                                                & South Korea & Moon, HT        & \begin{tabular}[c]{@{}l@{}}Korea Advance Institute of \\ Science \& Technology\end{tabular} & 12 & 0.41 \\ \hline
\textbf{19}                                                & China                            & Xiong, X      & Tianjin University                                                                       & 17 & 0.40 & \textbf{44}                                                & Germany     & Mimkes, J       & \begin{tabular}[c]{@{}l@{}}University of Gesamthsch \\ Paderborn\end{tabular}               & 11 & 0.36 \\ \hline
\textbf{20}                                                & Japan                            & Takahashi, T  & University of Tokyo                                                                      & 17 & 0.15 & \textbf{45}                                                & South Korea & Kim, S          & \begin{tabular}[c]{@{}l@{}}Korea Advance Institute of \\ Science \& Technology\end{tabular} & 11 & 0.29 \\ \hline
\textbf{21}                                                & China                            & Zhong, LX     & Dianzi University                                                                        & 16 & 0.33 & \textbf{46}                                                & South Korea & Oh, G           & \begin{tabular}[c]{@{}l@{}}Pohang University of \\ Science \& Technology\end{tabular}       & 11 & 0.31 \\ \hline
\textbf{22}                                                & China                            & Chen, W       & Shenzhen Stock Exchange                                                                  & 16 & 0.42 & \textbf{47}                                                & USA         & Amaral, LAN     & Northwestern University                                                                     & 11 & 0.46 \\ \hline
\textbf{23}                                                & England                          & Scalas, E     & University of Sussex                                                                     & 16 & 0.15 & \textbf{48}                                                & China       & Yang, G         & Fudan University                                                                            & 10 & 0.49 \\ \hline
\textbf{24}                                                & Italy                            & Lillo, F      & Scuola Normale Super Pisa                                                                & 16 & 0.37 & \textbf{49}                                                & China       & Zhang, YJ       & Tianjin University                                                                          & 10 & 0.51 \\ \hline
\textbf{25}                                                & Japan                            & Fujiwara, Y   & University of Hyogo                                                                      & 16 & 0.30 & \textbf{50}                                                & India      & Chakraborti, A  & Jawaharlal Nehru University                                                              & 10 & 0.29 \\ \hline
\end{tabular}%
}
\label{table:50Aauthors}
\end{table}

\begin{table}[]
\caption{List of 50 institutes based on the degree (collaboration). The table shows the institute name, number of collaborations ($k$) and number of authors (\#a)}.
\begin{tabular}{|l|l|l|l|l|l|l|l|}
\hline
\textbf{S.No.} & Institutes                                                                                               & $k$  & \#a & \textbf{S.No.} & Institutes                                                                                           & $k$  & \#a \\ \hline
\textbf{1}     & Boston University,   USA                                                                                 & 56 & 30  & \textbf{26}    & University of Evora, Portugal                                                                        & 14 & 3   \\ \hline
\textbf{2}     & \begin{tabular}[c]{@{}l@{}}East China University of \\ Science \& Technology, China\end{tabular}         & 36 & 38  & \textbf{27}    & CNRS, France                                                                                         & 13 & 3   \\ \hline
\textbf{3}     & University of Palermo, Italy                                                                             & 34 & 13  & \textbf{28}    & \begin{tabular}[c]{@{}l@{}}Saha Institute of Nuclear \\ Physics,  India\end{tabular}                 & 13 & 10  \\ \hline
\textbf{4}     & University Buenos Aires,  Argentina                                                                      & 26 & 5   & \textbf{29}    & Sony Compter Science Labs,  Japan                                                                    & 13 & 2   \\ \hline
\textbf{5}     & University of Tokyo,  Japan                                                                              & 25 & 17  & \textbf{30}    & \begin{tabular}[c]{@{}l@{}}Swiss Federal Institute of \\ Technology,  Switzerland\end{tabular}       & 13 & 6   \\ \hline
\textbf{6}     & Int Christian University,  Japan                                                                         & 23 & 2   & \textbf{31}    & Trinity College   Dublin,  Ireland                                                                   & 13 & 6   \\ \hline
\textbf{7}     & University of Leicester,  England                                                                        & 23 & 8   & \textbf{32}    & \begin{tabular}[c]{@{}l@{}}Federal University of   \\ Rio Grande do Sul, Brazil\end{tabular}         & 13 & 6   \\ \hline
\textbf{8}     & Santa Fe Institute, USA                                                                                  & 22 & 9   & \textbf{33}    & University of Pavia, Italy                                                                           & 13 & 3   \\ \hline
\textbf{9}     & UCL,  England                                                                                            & 21 & 8   & \textbf{34}    & \begin{tabular}[c]{@{}l@{}}Artemis Capital Asset   \\ Management GmbH, Germany\end{tabular}          & 12 & 1   \\ \hline
\textbf{10}    & Kyoto University, Japan                                                                                  & 19 & 7   & \textbf{35}    & Kings College London, England                                                                        & 12 & 10  \\ \hline
\textbf{11}    & Tokyo Institute of   Technology,  Japan                                                                  & 19 & 13  & \textbf{36}    & Korea   University,  South Korea                                                                     & 12 & 3   \\ \hline
\textbf{12}    & Aalto   University,  Finland                                                                             & 18 & 7   & \textbf{37}    & Peking University, China                                                                             & 12 & 9   \\ \hline
\textbf{13}    & Ist Nazl Fis   Nucl,  Italy                                                                              & 18 & 2   & \textbf{38}    & Tel Aviv University, Israel                                                                          & 12 & 6   \\ \hline
\textbf{14}    & \begin{tabular}[c]{@{}l@{}}Korea Advance Institute of \\ Science \& Technology, South Korea\end{tabular} & 18 & 11  & \textbf{39}    & University of Catolica Brasilia, Brazil                                                              & 12 & 3   \\ \hline
\textbf{15}    & University of Wroclaw,  Poland                                                                           & 18 & 7   & \textbf{40}    & \begin{tabular}[c]{@{}l@{}}University of Electronic \\ Science and Technology, China\end{tabular}    & 12 & 7   \\ \hline
\textbf{16}    & University Maryland, USA                                                                                 & 17 & 5   & \textbf{41}    & University of Fed Alagoas, Brazil                                                                    & 12 & 8   \\ \hline
\textbf{17}    & Bar Ilan University, Israel                                                                              & 16 & 4   & \textbf{42}    & University of Kiel, Germany                                                                          & 12 & 3   \\ \hline
\textbf{18}    & University Cologne, Germany                                                                              & 16 & 6   & \textbf{43}    & University of Piemonte Orientale, Italy                                                              & 12 & 2   \\ \hline
\textbf{19}    & Kanazawa Gakuin University, Japan                                                                        & 15 & 2   & \textbf{44}    & University of Politecn Madrid, Spain                                                                 & 12 & 6   \\ \hline
\textbf{20}    & National University of Singapore                                                                         & 15 & 5   & \textbf{45}    & University of Porto, Portugal                                                                        & 12 & 6   \\ \hline
\textbf{21}    & \begin{tabular}[c]{@{}l@{}}University of California \\ Los Angeles, USA\end{tabular}                     & 15 & 1   & \textbf{46}    & \begin{tabular}[c]{@{}l@{}}Budapest University of \\ Technology \& Economics, Hungary\end{tabular}   & 11 & 7   \\ \hline
\textbf{22}    & University Politecn Marche, Italy                                                                        & 15 & 2   & \textbf{47}    & \begin{tabular}[c]{@{}l@{}}Pohang University of   \\ Science \& Technology, South Korea\end{tabular} & 11 & 5   \\ \hline
\textbf{23}    & \begin{tabular}[c]{@{}l@{}}Complexity Science Hub \\ Vienna,  Austria\end{tabular}                       & 14 & 3   & \textbf{48}    & University of Adelaide, Australia                                                                    & 11 & 2   \\ \hline
\textbf{24}    & \begin{tabular}[c]{@{}l@{}}Consejo Nacl Invest   \\ Cient \& Tecn, Argentina\end{tabular}                & 14 & 3   & \textbf{49}    & University of Liege, Belgium                                                                         & 11 & 1   \\ \hline
\textbf{25}    & ETH, Switzerland                                                                                         & 14 & 6   & \textbf{50}    & Zhejiang University, China                                                                           & 11 & 27  \\ \hline
\end{tabular}
\label{table:50Uni}
\end{table}

\section{Discussion and conclusion}
\label{sec:summary}

We have presented the detailed analysis of ``Econophysics'' in terms of the evolution and structure of collaborations networks from 2000-2019.
We have performed a systematic empirical research highlighting the patterns in data, key disciplines by cited references, and the patterns of collaborations at micro, meso, and macro-levels.
The key findings of the study are: (i) The impact of self-citations on citations reveals that in first few years the publications have received more self-citations and this trend goes down with time. Also, on an average a paper has received first self-citations in first two years after the publication. (ii) The disciplines extracted from cited references from all published papers highlights the higher contribution of \textit{physics} and second highest of \textit{economics}. The higher contribution of physicists' towards the growth of Econophysics reveals the true nature of the discipline. (iii) The co-authorship network at micro-level identifies the key authors and their contributions as an individual or in group.
Also, the number of papers contributed by teams of varying sizes and the evolution of the team size over time is presented. We identified communities inside the giant component of the network and presented the relationships among nodes degrees and centrality measures (betweenness, closeness and eigencentrality). (iv) We also explored the authors' affiliations and country collaborations at meso and macro level. Results highlight that large number of authors are affiliated to a few numbers of institutions and China and USA has produced the higher authors as well as institutions. In terms of national and international collaborations, China share more national and USA shares more international collaborations. (v) Finally, the author's collaborations and affiliations networks are explored in terms of average degree, average clustering coefficient, average path length, size of giant component, etc. to study the networks evolution with a yearly resolution.

To conclude further, our study has provided an integrated view of citation dynamics and the growth of scientific collaborations networks of Econophysics metadata from 2000-2019. Our study justified the highest contribution of physicists' towards the field and to spread the visibility of the discipline, we suggest authors should publish more in interdisciplinary journals.
However, the low number of publications reported under the Econophysics domain in Web of Science points out as a limitation of the study which further leads to the absence of the significant contribution of few authors.
A possible future direction to extend the study is to integrate temporal data and quantify the evolution process of the co-authorship network and affiliations network~\citep{borner2004simultaneous}. This could reveal how the importance of an author varies with time at different stages in his/her career.

\section*{Acknowledgment}
This work greatly benefited from discussions with and comments from A. Chakraborti and H. K. Pharasi. This work uses Web of Science data by Thomson Reuters provided by the Northwestern University.
\printcredits

\bibliographystyle{cas-model2-names}

\bibliography{cas-refs}

\end{document}